# Mining Bad Credit Card Accounts from OLAP and OLTP

Sheikh Rabiul Islam
Tennessee Technological University
1 William L Jones Dr
Cookeville, TN 38505
+1 (931) 372-3101
sislam42@students.tntech.edu

William Eberle
Tennessee Technological University
1 William L Jones Dr
Cookeville, TN 38505
+1 (931) 372-3101
weberle@tntech.edu

Sheikh Khaled Ghafoor
Tennessee Technological University
1 William L Jones Dr
Cookeville, TN 38505
+1 (931) 372-3101
sghafoor@tntech.edu

## ABSTRACT
Credit card companies classify accounts as a good or bad based on historical data where a bad account may default on payments in the near future. If an account is classified as a bad account, then further action can be taken to investigate the actual nature of the account and take preventive actions. In addition, marking an account as "good" when it is actually bad, could lead to loss of revenue – and marking an account as "bad" when it is actually good, could lead to loss of business. However, detecting bad credit card accounts in real time from Online Transaction Processing (OLTP) data is challenging due to the volume of data needed to be processed to compute the risk factor. We propose an approach which precomputes and maintains the risk probability of an account based on historical transactions data from offline data or data from a data warehouse. Furthermore, using the most recent OLTP transactional data, risk probability is calculated for the latest transaction and combined with the previously computed risk probability from the data warehouse. If accumulated risk probability crosses a predefined threshold, then the account is treated as a bad account and is flagged for manual verification.

## CCS Concepts
• **Information systems➝ Information systems applications**  • **Data mining➝ Collaborative filtering**

## Keywords
OLTP, Data warehouse, Risk Probability, Classifier, WEKA.

## 1. INTRODUCTION
Credit cards are usually issued by a bank, business or other financial institution that allows the holder to purchase goods and services on credit. A person can have multiple credit cards from different companies. Companies who provide credit scores suggest card holders use multiple credit cards in order to increase their credit score. A credit score is a three-digit number between 300 and 850 that indicate the creditworthiness of a person. The credit score is used by lenders to determine someone's credit worthiness for various lending purposes.

A credit score can affect whether or not someone is approved for credit as well as what interest rate they will be charged [6]. Recklessly using multiple credits card is one of the reasons that someone is unable to pay their credit card bill on time, which can eventually turn into long-term debt for the card holder. Other reasons for being unable to pay their bill include job loss, health issues, or an inability to work, which can eventually result in "bankruptcy ". In any case, this becomes an issue for both the credit card companies and the customer.

To address this problem, besides carefully evaluating the creditworthiness of credit card applicants at the very beginning, the credit card issuer needs to identify potential bad accounts that are at the risk of going to bankruptcy over the life of their credit. From the creditor's side, the earlier the bad accounts are identified, the lower the losses [7]. A system that can identify these risky accounts in advance would help credit card companies to take preventive actions. They could also potentially communicate information to the account holder and provide suggestions for avoiding bankruptcy.

Online Analytical Processing (OLAP) systems typically use archived historical data from a data warehouse to gather business intelligence for decision-making.  On the other hand, Online Transaction Processing (OLTP) systems, only analyze records within a short window of recent activities - enough to successfully meet the requirement of current transactions [8]. Older transactional data are usually excluded from OLTP due to performance requirements and usually archived in the data warehouse. To compute the risk factor associated with an account both historical transactional data and recent transactions should be used to get a more accurate picture. In this paper, we propose a framework that computes the risk factor of a credit card account using both archived data from the data warehouse *as well as* recent transactions from OLTP. In our framework, the risk probability from the recent transactions is calculated using two methods: *Standard Transaction Testing* along with *Adaptive Testing*. We have validated our framework using a dataset from a German credit company found publicly on the internet [5].  Our approach can be used to predict whether an account is bad or good in real time as a transaction occurs. Our approach can then be used by a credit card company to take a more proactive action when it comes to verifying transactions and a customer's ability to pay.

The following section contains some of the related work. The framework for our approach is described in section 3. Section 4 describes the data sources with data preparation and feature extraction. Experiments, Standard Transaction Testing, and Adaptive Testing are described in Section 5. Section 6 contains results, and section 7 presents our conclusions and future work

## 2. RELATED WORK

A significant amount of research has been done in the area of financial fraud analysis and detection, especially in credit card fraud detection. However, not much work has been done in the area of working with transactional data from an OLTP system to predict bad credit card accounts. The research work of [1] is a dynamic model and mechanism to discover fraud detection system limitations while existing fraud detections systems use some predefined rules and scenarios or static models. In this instance, their dynamic model updates rules periodically [1]. They use a KDA clustering model which is a combination of three clustering algorithms, k-means, DBSCAN and the Agglomerative clustering algorithm, that are then represented together as a dynamic solution [1]. However, with this approach, the accuracy obtained by KDA modeling for online data is much less than that of the offline data.

In the work of [3], the authors discuss different methods on fraud detection based on decision trees using gini impurity, information gain and a binary decision diagram. In the work presented in [4], a data mining approach is presented using transaction patterns for credit card fraud detection where the spending pattern may change anytime due to changes in income and preferences.

In summary, most work targets credit card fraud detection and is performed on offline data from a data warehouse. Our proposed approach is novel in that it works on data from *both* OLTP and OLAP systems.

## 3. PROPOSED FRAMEWORK

Our proposed framework computes the risk factor of an account by combining the risk probability from archived data in a data warehouse with the risk probability of a current transaction from OLTP. The risk probability from the archived data or data warehouse is precomputed and is stored as summarized data. The risk probability from OLTP is computed in real time as transactions occur and combined with the precomputed risk to determine the overall risk factor. Figure 1 shows the high-level diagram of our proposed framework. Figure 2 shows the flowchart of our proposed framework. Whenever a new transaction occurs in the OLTP system, it is passed through a *Standard Transaction Testing* process that checks whether the transaction deviates from any of standard rules.

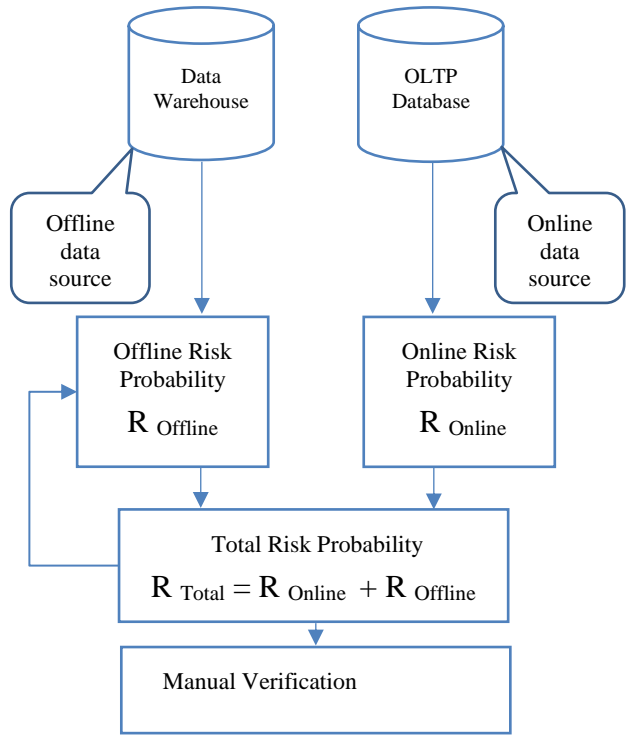

**Figure 1. High Level Diagram of the Proposed Framework.**

In the work of [7], the authors present a system to predict personal bankruptcy by mining credit card data. In their application, each original attribute is transformed either to a binary [good behavior and bad behavior] categorical attribute or multivalued ordinal [good behavior and graded bad behavior] attribute. Consequently, they obtain two types of sequences, i.e., binary sequences and ordinal sequences. Later they resort to a clustering technique for discovering useful patterns that can help them to identify bad accounts from good accounts. Their system performs well, however, they only use single data sources, whereas the bankruptcy prediction systems of credit bureaus use multiple data sources related to creditworthiness.

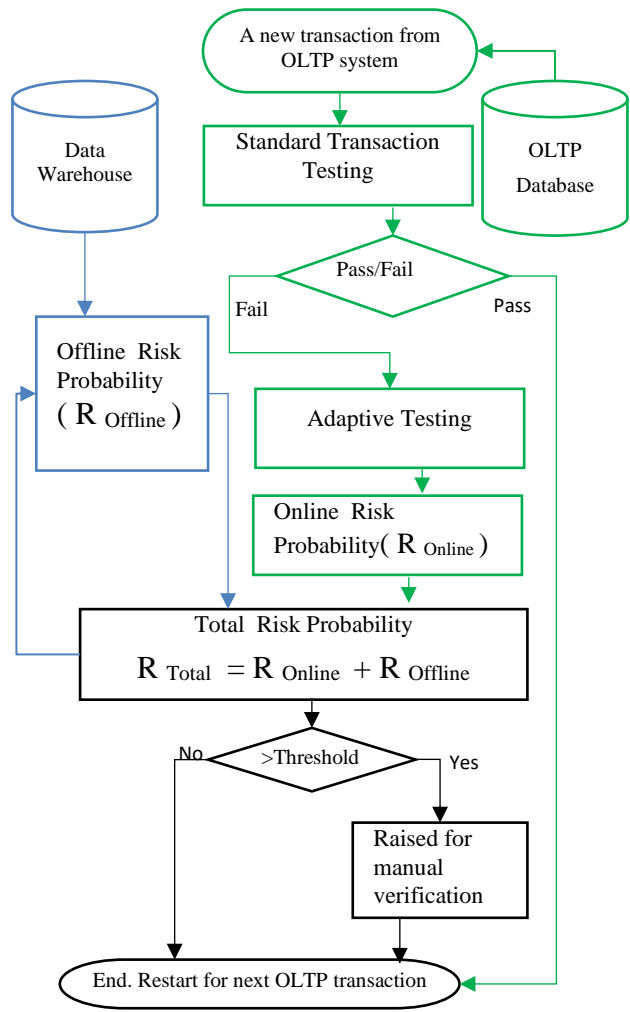

**Figure 2. Proposed Framework Flowchart.**

If the transaction passes the *Standard Transaction Testing*, then no further testing is done and the system continues with the next transaction. However, if the transaction fails the Standard Transaction Testing, then the transaction is passed to the *Adaptive Testing* process where customer specific measures are taken into account to measure the deviation and the risk probability from online data. Calculating risk probability from offline data is asynchronous to calculating risk probability from online data. It is a one-time job and is done at the beginning while configuring the system for our framework. When a transaction occurs in the OLTP system, the combined risk probability is calculated for that transaction which is stored and contributes to the risk probability from offline data for the next transaction from the same account. When we created the initial framework, we experimented with various popular classifiers (e.g., Naïve Bayes, J48, Rotation Forest, etc.) on the offline data. From our initial experiments, we discovered that Random Forest returns the highest number of correctly classified instances, as well as having the fastest execution times. The result is a risk probability for the offline data. Detail comparison of the result from the classifiers is discussed in the result section.

We express risk probability from offline data as $R_{Offline}$ and the risk probability form online data as $R_{Online}$. Risk probability from online data and risk probability from offline data are combined to get the combined risk probability (online + offline). We then express our overall or total risk probability as $R_{Total}$, which is equivalent to $R_{Online} + R_{Offline}$. After subsequent transactions for the same account, the combined risk probability $R_{Total}$ of the current transaction is combined with the offline risk probability for the same account. If the combined risk probability is greater than the threshold then the account is flagged for manual verification, otherwise the process ends here and starts from the beginning for the next transaction. We use a combined risk probability interchangeably with an overall or total risk probability in rest of this paper.

## 4. DATA
### 4.1 Data Sources
Finding large and interesting sources of financial data is challenging as these data are not made available to the research community because of obvious privacy issues. In this work, we use a dataset of a German credit company found publicly on the internet for research purposes [5]. That data contains both a credit summary, as well some anonymized detail information. There is summarized account information of 1000 accounts with 24 features or attributes. This is a labeled dataset where each account is labeled as good or bad (1 or 0). Table 1 shows what were determined to be the important features – a process that will be defined in the next section.

OLTP data consists of actual, real-time, transaction data. Unfortunately, a real transaction dataset that directly corresponds to this offline (summary or profile) dataset is not currently available to the research community. Therefore, in order to present a proof-of-concept, we will list some of the real OLTP transactions taken from the credit cards online portal [12], shown in Table 3, along with example use cases, shown in Section 5, that demonstrate how they might contribute to the risk possibility calculation.

### 4.2 Offline Data Preparation
In order to process the data and apply our models, we will use the publicly available machine learning tool WEKA [11]. WEKA requires the input dataset in a format called ARFF. An ARFF file is an ASCII text file that describes a list of instances sharing a set of attributes. ARFF files have two distinct sections. The first section is the header information that contains the relation names, a list of attributes (the columns in the data), and their types. The second section contains the data information [10]. We will use an online conversion tool [9] to convert the offline data (account summary and profile) into the ARFF format. We will then use the ARFF file as input to WEKA for our experiments.

**Table 1. Attributes from German credit dataset**

| Attribute | Type | Example value |
|---|---|---|
| Status of existing checking account | Qualitative | No checking accounts, salary assignment for at least 1 year, >=1000 |
| Duration in month | Numerical | 12 months |
| Credit history | Qualitative | No credit taken, all credit paid duly, |
| Present employment since | Qualitative | <1 year, < 4 years |
| Personal status | Qualitative | Male: divorced/separated, Female: single/married |
| Present residence since | Numerical | 24 months |
| Age in years | Numerical | 28 years |
| Housing | Qualitative | Rent, own |
| Job | Qualitative | Skilled employee, self-employed |
| Foreign Worker | Qualitative | Yes, no |

### 4.3 Feature Extraction
In order to reduce the possibility of over-fitting in our classifier, we need to minimize the number of features or attributes that the model uses, keeping only those that are the most informative. In order to determine which attributes are important and informative, we use the *feature selection with filter* option in the WEKA data mining tool that uses an attribute evaluator and a ranker to rank all features in the dataset. For the attribute evaluator, we use "InfoGainAttributeEval" that evaluates the worth of an attribute by measuring the information gain with respect to the class. The "Ranker" ranks each attribute by their individual evaluation in conjunction with the attribute evaluator "InfoGainAttributeEval". This is a supervised approach using the ARFF file mentioned in the previous section, to train the model. Figure 3 shows a portion of the output that we get using the attribute selection method in the WEKA data mining tool. It ranks all attributes from our offline data source and assigns a ranked value for each attribute. We then discard those attributes that will have no effect on the result of classification (i.e., those with a rank value of zero). The greater the value of the rank the more important the attribute. That means if we exclude an attribute which has a high rank then the accuracy of the classification will drop sharply. We use this ranked value to calculate the impact factor of an attribute in our experiments.

```
Attribute selection output

Attribute Evaluator (supervised, Class (nominal):
      Information Gain Ranking Filter
Ranked          attributes:
0.094739        statusofexistingcheckingaccount
0.043618        credithistory
0.0329          durationinmonth
0.013102        presentemploymentsince
0.005823        foreignworker
0               presentresidencesince
```

**Figure 3. Attribute selection using WEKA Attribute selector.**

## 5. EXPERIMENTS

For the purpose of calculating the risk probability from offline data, we set up a classification experiment that gives the risk probability for each account based on offline summary data. As mentioned previously, we already have summarized information for 1000 accounts with their credit history. Among those, we use 50% of these accounts to train the classifier and the remaining 50% to test the classifier. We are interested in the probability value of being a bad account or a good account for each of the accounts from the classifiers results. This is the risk probability from the offline data generated from the WEKA data mining tool. As mentioned previous, we selected the Random Forest classifier as it provides the best accuracy and execution time. So, using WEKA's Random Forest classifier, we calculate the risk probability of the offline data. The classification gives us the binary result of each account for being bad or good based on offline data. *Random Forest* classifier gives us best results with a highest rate of correctly classified instance of 74.6%. But we are not using that binary result from classification. Instead, we are interested in the *probability value* for each account being bad or good based on the offline data. Random Forest also takes the lowest execution time .01 second to classify provided 500 instances of test data.

We run all classifiers in WEKA 3.8.1 mostly using the default parameters for all of the attributes unless otherwise specified. The machine that we have used to do this experiment is a commodity machine with 6GB RAM, Intel Core i5 CPU with a speed of 1.8 GHz. For *Naïve Bayes*, we choose the base version where all default parameters were used. We used *J48* classifier with pruning. In the case of *Random Forest*, we set "*breakTiesRandomly*" to true to break ties randomly when several attributes look equally good. And for *Rotation Forest,* all default parameters were used. We then take the probability value from the *Random Forest* as the value of the risk probability from the offline data for each account. In Section 6, we show the procedure of retrieving the offline risk probability from the offline data. Figure 4 shows the probability distribution of being a bad or good account based on offline data for each of 500 accounts.

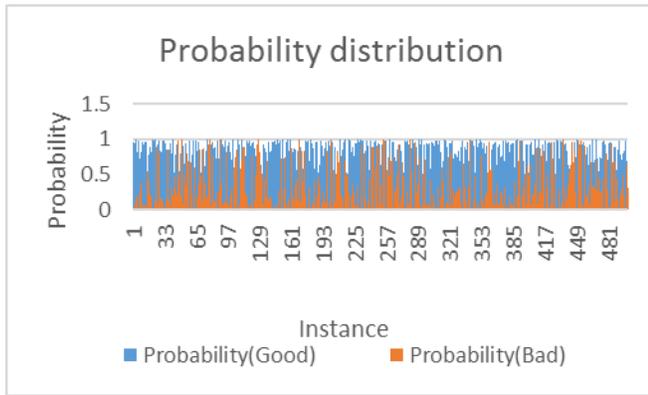

**Figure 4. Probability distribution of offline data classification.**

Figure 5 shows the density of probability instances. Here one interesting thing is evident and that is for most of the instances, the probability value tends to be in the upper or lower extremes rather than in middle, leading to a more accurate overall risk probability as values in the middle areas tend to show higher false positive rates. For the purpose of getting the risk probability from online data, we present our two methods: *Standard Transaction Testing* and *Adaptive Testing* (as shown in Figure 2 and discussed in Section 5). Remember, in order to be a real-time system, each new transaction from the OLTP system is passed through this framework as soon as the transaction occurs.

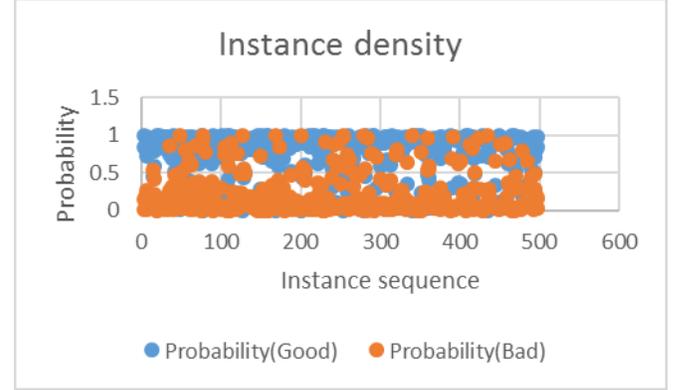

**Figure 5. Instance Density.**

### 5.1 Standard Transaction Testing

The purpose of this test is to identify transactions that deviate from the normal behavior and pass them to the next test named *Adaptive Testing*. For the *Standard Transaction Testing*, we have made a Standard Rule listing in Table 2. This is a partial collection of rules that every normal or good transaction require to follow according to our proposed framework. While this is just an initial set of rules based on the perception, it is possible to add as many as rules needed in this table based on future requirement. This *Standard Rules* table contains all the rules that reflect standard and normal behavior.

**Table 2. Standard Rules**

| Rule ID | Rule |
|---|---|
| 1 | Transaction amount $<= \Sigma ( \mu_{\text{transaction amount}} + \sigma_{\text{transaction amount}})$ |
| 2 | Number of transaction per day $<= \Sigma ( \mu_{\text{number of transaction}} + \sigma_{\text{number of transaction}} )$ |
| 3 | Payment within due date |
| 4 | Minimum amount due paid |
| 5 | Paid amount greater than or equal to due amount |
| 6 | Transaction location is near user's physical location |

The first rule we have in our table is, whether the transaction amount is less than or equal to the summation of the average transaction amount ($\mu_{\text{transaction amount}}$) and the standard deviation of the transaction amount ($\sigma_{\text{transaction amount}}$). The next standard rule regards whether the number of transaction per day for an account is less than or equal to the summation of the average number of transactions per day per account ($\mu_{\text{number of transaction}}$) and the standard deviation of the number of transactions per day per account ($\sigma_{\text{number of transaction}}$). This can help in identifying risky transactions. Other standard rules are included to indicate a common set of rules, and are self-explanatory. Each new transaction from the OLTP system is validated according to the standard rules defined in Table 2. The flexibility of our proposed approach allows for users to add as many standard rules as needed. In summary, the *Standard Rules* perform the primary screening of transactions.

We have collected some real OLTP credit card transactions to explain the OLTP risk probability calculation. We have anonymized the account number and transaction number for those transactions. Though accounts for these OLTP transactions have no direct mapping with the accounts of the offline data that we have. As we couldn't find a dataset comprises of both online and offline data for the same set of accounts. Still, we have selected these OLTP transactions very carefully to show a realistic relation among both online and offline data.

**Table 3. OLTP Transaction**

| TID | AC | Tran. Date | Description | Amount ($) | Category |
|---|---|---|---|---|---|
| 1 | 1 | 2017-01-20 | SOUTHWES5268506576536 800-435-9792 TX | 237.90 | Airlines |
| 2 | 2 | 2017-01-20 | INTERNET PAYMENT - THANK YOU | 25.00 | Payments and Credits |
| 3 | 3 | 2017-01-20 | DNH*GODADDY.COM 480-505-8855 AZDNH*GODADDY.COM | 155.88 | Merchandise |
| 4 | 4 | 2017-01-20 | WM SUPERCENTER #657 COOKEVILLE TN | 102.88 | Supermarkets |
| 5 | 5 | 2017-01-20 | BESTBUYCOM 775203010161 888-BESTBUY MN | 131.69 | Merchandise |

Here, TID = Transaction Id, AC = Account Number and Tran. Date= Transaction Date. From Table 3, we can see that, there is a transaction for account number (AC) 1 with transaction id (TID) 1. And it is a transaction of $237.90 for an air ticket purchase from Southwest airlines. As soon as the transaction occurrs, it is passed to the *Standard Transaction Testing*. All rules of *Standard Transaction Testing* are not applicable to all transactions. There is a relevance mapping table that contains which standard transaction is relevant to which type of transaction. Here the type of the transaction is determined by the category of the transaction. For the first OLTP transaction, the "Air ticket purchase of $237.90" relevancy mapping and satisfactory result is listed in the table (Table 4).

If any rows of the relevancy table (Table 4) have the value "Yes" in the "Relevancy" field for a transaction, it means that the transaction is *relevant* to the rule. In a similar fashion, if the value of the "Satisfy" field is "Yes", the transaction *satisfies* the rule. Now we check to see if rules for which the transaction under test is relevant (Relevancy=Yes) but doesn't satisfy (Satisfy=No) the rule. That means to search for rows in Table 4 those have the value "Yes" in "Relevancy" column but "No" in the satisfy column. If we can find any such row, then the transaction has failed to pass the *Standard Transaction Testing*. As we can see from the table (Table 4), row 1 and row 4 has the value "Yes" in the "Relevancy" field but "No" in the "Satisfy" field. Thus, in this example, the transaction has failed to pass the *Standard Transaction Testing* and will need to be forwarded to the next test, *Adaptive Testing*, with a reference that the transaction has failed to satisfy Rule ID 1 and 4 of the *Standard Rules* table (Table 2).

**Table 4. Relevancy Mapping**

| Rule ID | Rule | Relevancy | Satisfy |
|---|---|---|---|
| 1 | Transaction amount $<= \Sigma(\mu_{transaction\ amount} + \sigma_{transaction\ amount})$ | Yes | No |
| 2 | Number of transaction per day $<= \Sigma(\mu_{number\ of\ transaction} + \sigma_{number\ of\ transaction})$ | Yes | Yes |
| 3 | Payment within due date | No | NA |
| 4 | Minimum amount due paid | Yes | No |
| 5 | Paid amount greater than or equal to due amount | No | NA |
| 6 | Transaction location is near user's physical location | Yes | Yes |

## 5.2 Adaptive Testing

The *Adaptive Testing* process is a test that is more customer centric rather than the standard rules that are applicable for every account in the same way. It takes customer specific measures like foreign national, job change, address change, promotion, salary increase, etc. into consideration. This is a kind of test that recognizes possible causes for which a transaction is unable to satisfy a rule in the *Standard Rules*. Table 5 represents a listing of some of the possible causes for which a transaction may fail to follow the relevant standard rules in Table 2.

There are two new columns in Table 5: "Impact" and "Impact coefficient". Attributes found by the WEKA *Attribute Selector* tool from the offline data described in the feature extraction section gives us a ranked value (Fig. 3) for each attribute. Based on this ranked value, and based on the relationship between an attribute and related rules, we compute the impact associated with each rule/cause, as specified in the "Impact" column of the table (Table 5). For example, a foreign worker is an attribute in our offline data, and the causes "Out of the country" and "Air ticket purchase" of *Adaptive Testing* are related to this attribute. This *impact* assignment is completely company specific and customizable based on different analysis. And impact coefficient is the coefficient of the impact. Attributes those have more information gain, we are assigning more impact to the causes related to those attributes. Still a company may want to give more importance on some adaptive causes than others based on the business requirement. By default, this impact assignment is based on the rank of the attribute and the relation of adaptive cause with the attributes. Though companies have the provision to overwrite the default impacts.

Returning to our previous example of a transaction of $237.90 for the air ticket purchase by Account "1", the transaction fails to pass the *Standard Transaction Testing* due to two reasons 1) transaction amount was above the summation of average transaction amount and the standard deviation of the transaction amount, and 2) minimum due of last month was unpaid. The transaction is then passed to the *Adaptive Testing* component, along with the offending rules from Table 2 (i.e., Rule ID 1, 4). The *Adaptive Testing* component then checks its rules table for all rules that contain the

value 1 and/or 4 in its "Related Standard Rule" column (as shown in Table 5).

**Table 5. Adaptive Rules**

| Rule/ Cause ID | Rule/Cause | Related Standard Rule | Impact | Impact coefficient |
|---|---|---|---|---|
| 1 | Address change | 6 | 1x | 1 |
| 2 | Air ticket purchase | 1,2 | 1x | 1 |
| 3 | Job switch | 3 | 2x | 2 |
| 4 | Out of the country | 3,4,1,6 | 2x | 2 |
| 5 | Foreign Worker | 3 | 2x | 2 |

From Table 5, we can see that Rule/Cause ID 2 and 4 have the value 1 and/or 4 in their "Related Standard Rule" column, meaning that the rules in row 2 and 4 are possible causes of breaking rules 1 and 4 of the *Standard Rules* (Table 2). So, we have got two possible causes: 1) air ticket purchase, and 2) out of the country for breaking the rule. In this case, the customer bought the air ticket but was not out of the country. Now we will explain the risk probability associated with this example (and others).

## 6. RESULTS

To get the offline risk probability, we run the *Random Forest* classifier on the offline data and take the risk probability value associated with each of each account. To recall, *Random Forest* gave us the highest *correctly classified instances* of 74.6% with the lowest execution time of .01 seconds. Table 6 shows the result from the top 4 classifiers for our experiment in terms of CCI and execution time.

**Table 6. Classifier Results on Offline Data**

| Cls | CCI % | ICI % | Avg. TP Rate | Avg. FP Rate | Pr | Re | Time |
|---|---|---|---|---|---|---|---|
| Naïve Bayes | 73.8 | 26.2 | .738 | .43 | .723 | .762 | .02 |
| Random Forest | 74.6 | 25.4 | .75 | .48 | .733 | .75 | .01 |
| Rotation Forest | 73.8 | 26.2 | .738 | .466 | .721 | .738 | .02 |
| J48 | 69.4 | 30.6 | .694 | .476 | .676 | .694 | .01 |

Here,
- CCI = Correctly Classified Instances
- ICI = Incorrectly Classified Instances
- Avg. TP Rate = TP/P
- Avg. FP Rate = FP/N
- Cls = Classifier
- Pr = Precision
- Re = Recall

Recall tells us how good a test is at detecting the positives, and Precision tells us how many positively classified instances were relevant. The "probability distribution" column in Figure 6 shows the probability values of an account to be bad or good, which is the offline risk probability for a particular account at the beginning. This is a snapshot from WEKA's *Random Forest* classifier on our offline data. For instance, in row 4 of Figure 6, there is 87.9% probability that the account is good and there is 12.1% probability that the account is bad. Here, "instance #" is the account number and "probability distribution" is the more granular result for that account rather than just the binary result good or bad.

```
Time taken to build model: 0.31 seconds

=== Predictions on test split ===

inst#,    actual, predicted, error, probability distribution
   1       2:2       1:1       +    *0.831  0.169
   2       2:2       1:1       +    *0.876  0.124
   3       1:1       1:1             *0.544  0.456
   4       1:1       1:1             *0.879  0.121
   5       1:1       1:1             *0.727  0.273
   6       1:1       1:1             *0.962  0.038
   7       1:1       1:1             *0.624  0.376
   8       2:2       2:2              0.376 *0.624
```

**Figure 6. Probability distribution (from classification result) on offline data.**

As we said before that our online data and offline data are not actually correlated. For the purpose of calculating overall risk probability, we need to establish a correlation among them. For that purpose, we are picking 5 random accounts and their offline risk probability from offline data (Table 7) and then assigning account number 1 to 5 to make a correlation with our online data.

**Table 7. Offline (preprocessed) risk probability**

| Account Number | Risk Probability from classification (%) |
|---|---|
| 1 | 70 |
| 2 | 23 |
| 3 | 82 |
| 4 | 79 |
| 5 | 43 |

Total risk probability for a transaction comes from both online and offline data. So, the equation of total risk probability is as below:

$$R_{Total} = R_{Online} + R_{Offline} \qquad (1)$$

Here,

$R_{Total}$ = Overall risk probability from both online and offline data.

$R_{Online}$ = Risk probability from online data

$R_{Offline}$ = Risk probability from offline data

Furthermore, risk probability from online data and offline data may carry different weights. For example, giving 60% weight to offline data and 40% weight to online data might provide better mining results for a particular company. On the other hand, for another company, a different combination of offline vs online risk probability weight might be better. So, the modified version of (1) for a total risk probability calculation is:

$$R_{Total} = \lambda R_{Online} + (1-\lambda) R_{Offline} \quad (2)$$

Where λ is the risk factor.

For our experiments, we are assuming that a 70% weight from online data and 30% weight from offline data which is an established ratio by the long-term tuning of our system for better mining results. So, for our case λ =.7 and 1- λ =.3. Data from both sources are more or less important for the total risk analysis. And this ratio can be adjusted based on trend analysis.

We can get the risk probability from offline data (R $_{Offline}$) for corresponding accounts from Table 7, which is related to the instance risk probability distribution value of classification results. To calculate the risk probability from online data (R $_{Online}$), we have derived the following equation:

$$R_{Online} = [1 - \frac{\Sigma \text{ Impact Coefficient }(X)}{\Sigma \text{ Impact Coefficient }(Y)}] \times 100 \quad (3)$$

Where X = Relevant valid rules from the *Adaptive Rules* table (Table 5) and Y = Relevant valid or invalid rules from the *Adaptive Rules* table (Table 5)

In other words, X is the collection of rules from *Adaptive Rules* table (Table 5) where the "Related Standard Rule" column has the value of any of the rule ids that are passed from *Standard Transaction Testing* and are valid causes for breaking a standard rule; and Y is the collection of rules from the *Adaptive Rules* table (Table 5) where the "Related Standard Rule" column has the value of any of the rule ids that are passed from *Standard Transaction Testing* irrespective of whether it is valid cause or not a valid cause. If no rule/cause is found in *Adaptive Rules* table (Table 5) for a transaction that is passed to *Adaptive Testing*, then the values of X and Y become zero. Thus, the value of R $_{Online}$ from (3) becomes 100%, which means the customer has no customer specific reason in the *Adaptive Rule* table resulting from assigning the highest online risk probability possible for that transaction. If there were some customer specific reasons, R $_{Online}$ would reduce by some ratio based upon the number of customer specific causes/rules available and the number of causes/rules among them that are valid for that transaction.

Using the example presented earlier, customer with id 1 has bought an air ticket but the customer is not out of the country or state yet. Rule id 1 and rule id 4 from standard rule table (Table 2) were relevant to the transaction but not satisfied. That's why the transaction was passed to "Adaptive Testing" with a reference to rule id 1 and 4. In the *Adaptive Test*, from Table 5 it is found that the row with "Rule/ Cause ID" 2 and 4 have the value 1 and or 4 in the "Related Standard Rule" column. So, either of rules *out of the country* or *Air ticket purchase* from the adaptive rules table (Table 5) is the cause of breaking the standard rules 1 and 4 for the transaction we are explaining. That gives us:

Y ={ Out of the country, Air ticket purchase}

But the customer's most recent location, which is usually appended with the OLTP transaction description, says that the customer is not out of the country (yet). So actually, out of the country is not a valid reason for breaking the standard rules, though it is relevant.

Thus, with X ={ Air ticket purchase }, using the equation (3):

$$R_{Online} = [1 - \frac{\Sigma \text{ Impact Coefficient }(X)}{\Sigma \text{ Impact Coefficient }(Y)}] \times 100$$

$$= [1 - \frac{\Sigma \text{ Impact Coefficient (Air ticket purchase)}}{\Sigma \text{ Impact Coefficient (Air ticket purchase)} + \text{Impact Coefficient (Out of the country)}}] \times 100$$

$$= [1 - \frac{1}{1+2}] \times 100$$

$$= .67 \times 100$$

$$= 67$$

From Table 7, we can see that offline risk probability for account 1 is 70%

So, R $_{Offline}$ = 70.

Putting these values in equation (2) and applying risk factors(λ) we get the overall risk probability for account 1 after the transaction 1 is recorded in the OLTP system. Thus, the risk factors(λ) is .7 for our case.

$$R_{Total} = \lambda R_{Online} + (1-\lambda) R_{Offline}$$

$$= .7 \times 67 + .3 \times 70$$

$$= 67.9$$

So, for the transaction that we are explaining, there is a chance of 67.9% that this account is going to be a bad account.

For the proof-of-concept, we are assuming a *Minimum Total Risk Probability Threshold* of 60% is established beforehand (by the user) based on the analysis of historical data. This means that if the total or overall risk probability is above 60%, then that transaction will be treated as a risky transaction (along with the associated account). In this example, the *Overall Risk Probability* (R $_{Total}$) is 67.9% and that is above the threshold 60%, so the account for that air ticket purchase transaction (Transaction 1) is suspended and raised for manual verification to justify the actual nature of the account.

When the overall risk probability for a transaction is completed, the offline risk probability is adjusted based on the value of R $_{Total}$, which affects the offline risk probability value of the next transaction for the same account. By this way, offline risk probability for an account gradually increases if the customer repeats similar transactions that are passed to the adaptive test from the standard testing. The risk probability threshold of 60% is not the only thing to consider. Besides the above tests, we are also interested to know how much the transaction being analyzed deviates from the median value. This will give us an idea of the deviation intensity of the transaction in terms of its risk probability. There is a precomputed *median risk probability* for each type of transaction over some period of time from both sources of data that is calculated separately from both Online and Offline data. The overall median risk probability is calculated simply by adding the online and offline median risk probability. Table 8 below lists the process of calculating the deviation from the median. We compare calculated risk probability for the transaction under experiment with the median risk probability value of that type of transaction in the case of online, offline and overall. The difference between them is called the *gap*, as shown in Table 8.

**Table 8. Calculating Gap**

|  | Online | Offline | Overall |
| --- | --- | --- | --- |
| Median risk probability of similar transaction | M1 | M2 | M3 |
| Risk probability of Transaction under experiment | L1 | L2 | L3 |
| Gap | X=L1-M1 | Y=L2-M2 | Z=L3-M3 |

If any of either X, Y, Z from Table 8 is greater than zero, we signal the transaction for manual verification and postpone activity of that account.

Furthermore, we need to see the spike from the previous transaction for the transaction under experiment. This gives us an idea of the intensity of the spike of the risk probability. Table 9 shows the way of calculating spike.

**Table 9. Calculating Spike**

|  | Online | Offline | Overall |
|---|---|---|---|
| Risk probability of just previous transaction for the same account | P1 | P2 | P3 |
| Risk probability of Transaction under experiment | L1 | L2 | L3 |
| Spike | X=L1-P1 | Y=L2-P2 | Z=L3-P3 |

Again, as before, if any of X, Y, Z from above table is greater than zero, we signal the transaction for manual verification and postpone activity of that account.

To date, we have yet to find comparable work for comparison. In addition, it should be pointed out that our approach is very efficient in terms of its quick detection of bad accounts. The primary reason of this efficiency is that the computation of the offline risk probability ($R_{Offline}$) is performed asynchronously with the calculation of the risk probability from the OLTP data and updated only in two situations: (1) if there are changes in the summary data in the data warehouse for that account, and (2) if a recent transaction's overall risk probability calculation is completed for that account. Once the overall risk probability calculation for a transaction is completed for an account, then that calculated risk probability contributes to the risk probability from the offline data while calculating the risk probability for an upcoming transaction from the same account. While we have been unable to obtain actual offline and online data for the same accounts, we have been able to demonstrate a proof-of-concept using some actual OLTP transactions taken from a credit card's online portal [12].

## 7. CONCLUSION

In this paper, we have presented a novel approach of mining bad credit card accounts from offline data in combination with recent transactions from OLTP data. This approach is very efficient in terms of the quick detection of bad accounts due to the matter that the offline and online risk calculation processes are asynchronous. Moreover, this approach is very versatile as any number of rules can be defined, and the overall risk probability ratio or weight can be adjusted based on requirements or the historical analysis of data, allowing for extensibility to supporting a real-time fraud detection approach to support organizations existing fraud detection system. Our work can be extended to a recommendation system that will help both the customers and the company with necessary recommendations and warnings so as to avoid bankruptcy. We are faced with some limitations in our data collection and data analysis, especially for finding a public OLTP dataset that is directly correlated to an offline public dataset in terms of bankruptcy. The reasons for this are: (1) corporations do not like to reveal their techniques to others; and (2) large and interesting sources of data are not made available to the academic community [7]. Our future plan is to build a real-world, integrated system based on our proposed framework. Our plan also includes using *High Performance Computing* to make the computations as efficient as possible for real-time decision making.

## 8. REFERENCES

[1] M. Vadoodparast, P. A. R. Hamdan, and D. Hafiz, "Fraudulent Electronic Transaction Detection Using Dynamic KDA Model ," (IJCSIS) International Journal of Computer Science and Information Security, vol. 13, no. 2, Feb. 2015.

[2] V. Priyadharshini and G. A. Macriga, "An Efficient Data Mining for Credit Card Fraud Detection using Finger Print Recognition," International Journal of Advanced Computer Research, vol. 2, no. 7, Dec. 2012.

[3] A. N. Pathak, M. Sehgal, and D. Christopher, "A Study on Fraud Detection Based on Data Mining Using Decision Tree," IJCSI International Journal of Computer Science Issues, vol. 8, no. 3, May 2011.

[4] J. W., Yoon and C. C. Lee, "A data mining approach using transaction patterns for card fraud detection," Jun. 2013. [Online]. Available: arxiv.org/abs/1306.5547.

[5] "UCI machine learning repository: Data set,". [Online]. Available: https://archive.ics.uci.edu/ml/datasets/Statlog+(German+Credit+Data. Accessed: Feb. 20, 2016.

[6] "Credit Karma: What is a good credit score?," Credit Karma. [Online]. Available: https://www.creditkarma.com/faq/what-is-a-good-credit-score. Accessed: Feb. 20, 2016.

[7] T. Xiong, S. Wang, A. Mayers, and E. Monga, "Personal bankruptcy prediction by mining credit card data," Expert Systems with Applications, vol. 40, no. 2, pp. 665–676, Feb. 2013.

[8] "Introduction to data warehousing concepts," 2014. [Online]. Available: https://docs.oracle.com/database/121/DWHSG/concept.htm#DWHSG9289. Accessed: Mar. 28, 2016.

[9] "Online CSV to ARFF conversion tool,". [Online]. Available: http://slavnik.fe.uni-lj.si/markot/csv2arff/csv2arff.php. Accessed: Mar. 28, 2016.

[10] "Attribute-relation file format (ARFF),". [Online]. Available: http://www.cs.waikato.ac.nz/ml/weka/arff.html. Accessed: Mar. 28, 2016

[11] "Weka 3 - Data Mining With Open Source Machine Learning Software In Java". Cs.waikato.ac.nz. N.p., 2017. Web. 15 Mar. 2017.

[12] "Sign On – Citibank ,". [Online]. Available: https://online.citi.com/US/JPS/portal/Index.do?userType=BB . Accessed: Mar. 20, 2017.